\documentclass[12pt]{article}
\usepackage{amsmath,amssymb} 
\usepackage{textcomp}
\usepackage{ifxetex}
\ifxetex
\usepackage[cm-default,no-math]{fontspec} 
\usepackage{xltxtra} 
\usepackage{polyglossia} 
\setdefaultlanguage{english} 
\defaultfontfeatures{Mapping=tex-text} 
  \newfontfamily\russianfont{Charis SIL}
  \newfontfamily\englishfont{Charis SIL}
\else
\usepackage[T1,T2A]{fontenc}
\fi
\usepackage{color,graphicx,shortvrb,epsfig}
\usepackage{url}
\usepackage{wrapfig}
\usepackage{wasysym}
\usepackage[dvipdfmx, dvips, bookmarks, bookmarksnumbered, bookmarksopen,colorlinks=true,linkcolor=black]{hyperref}

\usepackage{hyphenat}

\begin{document}
\begin{center}
\textbf{\Large \noindent Isolated neutron stars and studies of their interiors}\\[0.5cm]
S.~B.~Popov\\[0.5cm]
\end{center}
\begin{tabular}{l}
\textit{\indent Sternberg Astronomical Institute, 119991, Moscow, Russia}\\
\textit{\indent e-mail: polar@sai.msu.ru}
\end{tabular}
\begin{abstract}
\indent In these lectures presented at Baikal summer school on physics of elementary particles and astrophysics 2011, I present a wide view of neutron star astrophysics with special attention paid to young isolated compact objects and studies of the properties of neutron star interiors using astronomical methods. 

\end{abstract}

\begin{tabular}{l}
\hbox{PACS: 97.60.Jd, 26.60.-c, 97.80.Jp}\hfill \\
\end{tabular}

\section{Introduction.}

 Astronomical objects and phenomena became the subject of interest for
physicists since the Universe serves as a lab with many possibilities to
study physical processes under conditions unavailable on the Earth.  Among
others, neutron stars (NSs) provide a unique opportunity to study several
extreme phenomena in a single type of sources.  ``Wild'' properties of
NSs include:
\begin{itemize}
\item Ultrastrong magnetic fields, sometimes orders of magnitude above the
Schwinger value ($B_\mathrm{Sh}=m^2 c^3/ e (h/2 \pi) \sim 4.4 \,
10^{13}$~G).  
\item The strongest surface gravity among all known objects: $GM/R \sim 0.2
c^2$.
\item Central density up to an order of magnitude above the nuclear one.
\item Superfluidity of neutrons and protons.  
\end{itemize}
These properties bring NSs to attention of specialists working in such
different fields as quantum chromodynamics, theory of gravity, quantum
electro\-dyna\-mics, solid and condense state physics.

In these lectures I briefly review different subjects, mainly related to
phenomenology of NSs.  Special attention is paid to recent discoveries
related to young isolated NSs.  Then I discuss how different astrophysical
approaches (measurements of global parameters -- mass and radius, -- and
studies of cooling behaviour) help to uncover properties of NSs interiors. 
However, we start with some rapid excursion into the history of NS
exploration which is now a quite mature subject.

\subsection{History of neutron star studies}

 Traditionally, the history of NSs is tracked back to the paper by Landau published in 1932 \cite{Landau32}. Indeed, several important points have been proposed in this seminal note, however, it would be over-optimistic to expect more insights on NSs, even from a genius, before the discovery of a neutron! (See an important discussion on controversies related to the legend of NS prediction by Landau in the introductory part of the book \cite{hpy2007}.) So, real story of NS exploration can be began from papers by Baade and Zwicky, all published in 1934, especially important is \cite{bz1934}. In this it was correctly predicted that NSs are formed in supernova (SN) explosions, consist mainly of neutrons, and are very compact and dense.  
 
 In the following thirty years appeared several important papers, related to internal properties of NSs. At first, one should mention the construction of the main equation for the description of internal structure of compact objects taking into account general relativistic effects. Now this formula is known as Tolman-Oppenheimer-Volkoff equation \cite{t1939, ov1939}. In late 50s and in 60s different researchers started to work on realistic models for the equation of state (EoS) of dense matter. However, significant progress was obtained later.

 In 60s several very important results have been obtained in the Soviet Union. To name just a few of them. In 1959 Migdal \cite{m1959} noted that superfluidity is possible inside NSs. Ginzburg in 1964 \cite{g1964} suggested that  flux conservation in collapsing stars can lead to formation of highly magnetized compact objects, and Kardashev, also in 1964 \cite{k1964}, proposed that such a collapsar with high field and rapid rotation can power a nebula formed after the SN explosion. 

 Progress in understanding the internal properties of NSs made possible a discussion of the thermal evolution of these objects. First important papers on the subject appeared in mid-60s \cite{tc1964, morton, chiu}. Then, the world was ready to discover NSs.

 Curiously, in mid-60s NSs were already discovered, but could not be recognized, yet. Not to mention the fact that the Crab pulsar potentially could be identified in old images of the nebula both in optics and in radio,  NSs have been discovered as accreting X-ray sources in binaries, namely as Sco X-1 source \cite{gur1962}. Unfortunately, the first bright X-ray sources detected by rocket experiments were not pulsating. So, it was non-trivial to prove the presence of NSs in them. Thus, NSs were finally serendipitously discovered as radio pulsars \cite{bell1968}, unrelated to a prediction by Pacini \cite{pac1967}, that magnetized rapidly rotating NSs can emit electro-magnetic waves.

The history of NS studies is well described in many place, for example in \cite{hpy2007}.

\section{Astrophysical manifestations of neutron stars}

 During last $\sim 40$ years our knowledge about NSs increased a lot thanks to rapid development of observational technique. Now these sources are studied in all electro-magnetic bands, and we hope for registration of gravitational wave signals from coalescing NSs in the near future. 

  Still, radio pulsars form the most numerous population of NSs.~\footnote{See the on-line ATNF catalogue \cite{atnf}.} And accreting NSs in X-ray binaries hold the second place in the list of the most numerous. However,  results of last $\sim 15$ years demonstrated that NSs can appear in many flavours. In the following I shall give a sketch of the modern data on young isolated NSs which shows great diversity in parameters and appearances.  

 A NS can emit due to several different sources of energy. The first one is
rotation.  NSs can spin as fast as 1 msec.  With the moment of inertia
$I\sim 10^{45}$~g~cm$^2$ this corresponds to $E_{\mathrm{rot}}~\sim
10^{53}$~erg.  With a solar luminosity $L_\odot=2 \, 10^{33}$~erg~s$^{-1}$
this is enough for $10^{12}$~yrs, larger than the age of the universe $\sim
13-14$ billion years (but, of course, energy is never released in a uniform
way).  NSs can spin rapidly due to two reasons.  At first, most of them are
born with short spin periods due to significant contraction of initially
rotating core.  Then, a NS can be spun-up in a close binary system by
accretion, as the incoming matter brings angular momentum due to orbital
rotation.  Radio pulsars are examples of NSs emitting their rotational
energy.  So-called ``normal'' pulsars  are young objects, and
millisecond pulsars are objects ``recycled'' in interacting binaries.

 The second source of energy is -- accretion. As a NS is very compact,
falling of 1 gram of matter on its surface results in release of $\sim
10^{20}$~ergs -- more efficient that thermonuclear reactions!  Matter can come from a companion in a binary system, or from a debris disc, or from  interstellar medium.

 Then comes the residual heat. NSs are born in SN explosions, and have very high initial temperature, about 10 MeV. Most part of this energy is quickly carried out by neutrinos, and at the age $\sim 1-10$~yrs temperature of the bulk of a NS is $\sim 10^9$~K.  Thermal energy stored in a NS can reach $E_{\mathrm{therm}}\sim10^{48} T_9^2$~erg. This is enough to power a relatively bright source for several million years. The main part of a NS is isothermal with temperature during the first hundred thousands years $\sim 10^8$~K. But the external layers work as a heat insulator, so the surface temperature is lower, $\sim 10^6$~K. Without detailed calculation it is easy to estimate that a 10 km sphere with such temperature has a luminosity $L\sim 10^{32}$~erg~s$^{-1}$.  More detailed data will be given below. Now tens of young NSs are observed due their thermal residual emission.

 What remains as an energy source? Some heat can be released due to friction
in NS interiors.  Some can result from cracking of the crust due to sudden
stress relaxation.  But all this cannot contribute a lot in the total
budget.  However, there is one important source -- the magnetic field.

 NS magnetic fields can be very high (maybe for somebody it is easier to think about strong currents in NS interiors, and so about release of their energy). A NS cannot support fields above $\sim 10^{19}$~G. And such huge values, probably, are never reached. But fields $\sim 10^{15}$~G seem to be realistic. Then, the energy is $E_{\mathrm{mag}}\sim (B^2/8\pi)(4/3 \, \pi R^3)\sim B_{15}^2 \, 10^{47}$~erg. 
Sources which predominantly release their magnetic energy are called magnetars.

Now we briefly describe properties of several main types of  known  NSs. Most of the discussed objects are observed as X-ray sources, and I recommend \cite{mer-x} as a recent review on X-ray observations of young isolated NSs.

\subsection{Radio pulsars}

 Despite their name these objects are observed in all wavelengths. About
1600 ``normal'' and 300 recycled radio pulsars are known \cite{atnf}. 
Mostly they have been discovered in surveys of the galactic plane in the
radio waveband, and then many of them were identified in other spectral
ranges.  Several sources, including the famous Crab pulsar, are detected as
pulsating sources from gamma to radio.  Recently, the situation changed, and
many sources have been found by the gamma-ray observatory Fermi
\cite{fermi}.  Later, radio counterparts were spotted for some of them, but
not for all, which is due to the geometrical effects (gamma-ray beam is
often wider than the radio one, and they do not exactly coincide in
direction in many cases).

\begin{figure}[h]
 \centerline{\includegraphics[width=0.6\textwidth]{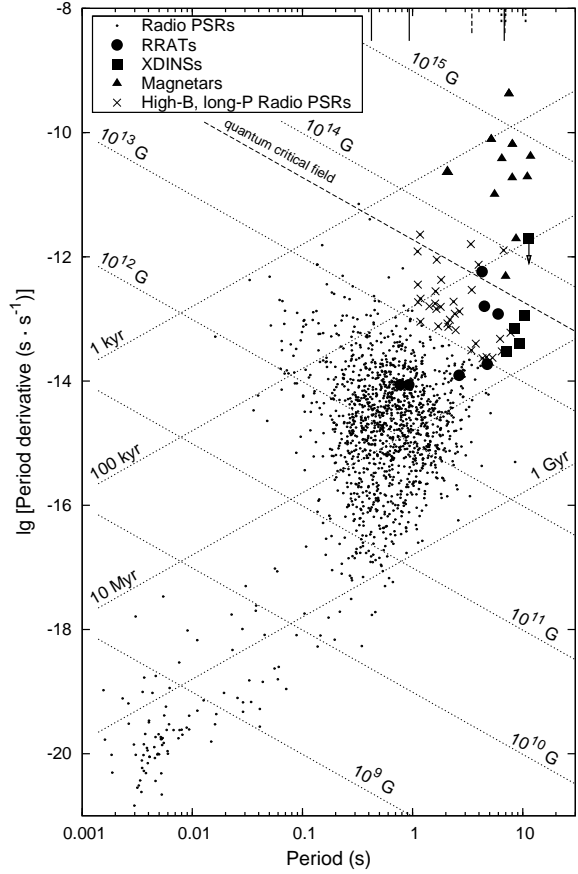}}
  \caption{ $P$~--$\dot P$ (spin period -- period derivative) diagram for radio pulsars and other sources. From \cite{kond}.}
\label{fig:ppdot}
\end{figure}

 Sources have spin periods in the range 0.001-10 sec (Fig. \ref{fig:ppdot}).
This parameter decreases as the sources evolve due to losses of the
rotational energy.  Only a tiny part of rotational energy is emitted in
radio.  Mostly, the energy is carried away by relativistic particle wind. 
The exact mechanism of energy losses and pulsar emission is still under
debate \cite{beskin}.  
However, a good approximation for the spin-down rate is given by a so-called
magneto-dipole formula:

\begin{equation}
\dot P= \frac{8\pi^2}{3Ic^3} \frac{B^2R^6}{P} \sin ^2 \chi 
\end{equation} 
Here, $P$ -- is the spin period, $B$ -- equatorial magnetic field, $I$ -- moment of inertia, $R$ -- radius of a NS, $\chi$ -- is the angle between spin and magnetic axis. 

The rotation energy losses in this approximation are:

\begin{equation}
L_{\mathrm{rot}}= \frac{2 (2 \pi)^4}{3 c^3} B^2 R^6/P^4\propto \dot P/P^3.
\end{equation}

These equations are typically used to estimate parameters of radio pulsars using known $P$ and $\dot P$.

 Recent review on radio pulsars can be found in \cite{beskin}. 

\subsection{Rotating radio transients}

 Progress in radio astronomy and data mining allowed detection of peculiar sources. In the radio waveband  it is  particularly difficult to identify very short flares from unknown objects, especially if they do not repeat often. Still, in 2006 it was announced about an important discovery \cite{rrats}. 

 Eleven sources producing very short --- 2-30 msec --- bursts have been
detected.  They are called RRATs --- Rotating RAdio Transients.  Bursts
ap\-pear\-ed to be produced by periodic sources with many ``empty''
cycles.  Periods are between 0.7 and 7 seconds, but typical intervals
between bursts are from several minutes up to few hours, which means that
mostly sources are not active (or below detectability level).  A period for
a given source is defined as the least common denominator for intervals
between individual dispersed pulses.  For several objects period derivatives
have been also measured, and obvious association with NSs was established.

In the $P$~--~$\dot P$ diagram the new sources occupy the region partly coinciding with the main radio pulsar cloud, partly extending towards longer spin periods and larger magnetic fields. On average, RRATs have larger magnetic fields and spin periods than standard radio pulsars. Initially, it was suspected that RRATs can represent a new distinct population of NSs. However, now it is usually accepted that  mostly RRATs do not form a separate class, but just demonstrate some specific type of activity in radio.

Further observations provided more --- several  dozens --- sources  of these
kind \cite{rrats_new}.  In addition, other types of activity were detected
from some of them.  In particular, more or less standard radio pulsar
emission was observed in many cases, and oppositely, some radio pulsars
demonstrate short intensive bursts, so if they were observed from larger
distance --- they would be classified as RRATs.  As a class RRATs are
identified as: ``.  .  .  repeating radio sources, with underlying
periodicity, which is more significantly detectable via its single pulses
than in periodicity searches'' \cite{rrats_new}.

 The most studied RRAT J1819-1458, as often happens, seems to be peculiar (see, for example \cite{rrats_new}). In the first place, it is necessary to say that this is the only of these sources detected not only in the radio waveband. It was identified as an X-ray source  in archival Chandra data \cite{reynolds2006}. 
Then it was observed several times in X-rays. Clear pulsations with the period coincident with the one determined in radio were detected. The spectrum appears to be thermal. In this respect the source looks similar to the Magnificent seven objects (see below). Peculiar glitches were reported from this source \cite{Lyne2009}. Also it is worth mentioning that this is a RRAT with the brightest bursts.

 Despite many attempts no other RRATs have been detected in any wave\-band, except radio. The nature of short radio bursts remains unclear, and we do not provide a review, or even a list, of models for this phenomena. 

\subsection{Central compact objects in supernova remnants}

 The history of this type of sources (CCOs for short) can be started with
X-ray observations with the HEAO-2 (Einstein) satellite.  Good angular
resolution of this instrument allowed to identify point sources in some
supernova remnants (SNRs).  In 1980 a central compact objects in the SNR
RCW103 was dis\-co\-ver\-ed \cite{rcw103}.  Few years later Tuohy et al. 
showed that the spectrum is consistent with the blackbody emission.  That
was drastically different from what was known and expected.  The pulsar in
the Crab nebula --- a classical prototype of all young NSs at that time ---
showed the spectrum dominated by non-thermal emission.  One by one more CCOs
which appeared very different from classical radio pulsars have been
discovered.  And at the end of the 20th century it became clear, that
despite small number of known sources of this kind their birth rate can make
a significant fraction of the total NS birth rate.  In the abstract of
\cite{gv99} it was written: ``Remarkably, these objects, taken together,
represent at least half of the confirmed pulsars in supernova remnants. 
This being the case, these pulsars must be the progenitors of a vast
population of previously unrecognized neutron stars.'' So, a clear
understanding of the existence of a distinct population of young NSs was
formed.

 Now up to ten CCOs are known \cite{cco}. For some of them spin periods and
period derivatives a measured.  In all cases small values for dipolar
magnetic fields (responsible for the spin-down) have been reported.  Fields
are about 1-2 orders of magnitude smaller than in normal radio pulsars, so
these sources are dubbed ``anti-magnetars'' \cite{antimag}.

\subsection{The Magnificent seven \& Co.}

 In early 70s it was proposed that isolated NSs can come to the stage of
accretion from the interstellar medium \cite{sh70, ost70}.  Such objects
were expected to appear as dim ($L\sim 10^{30}$~erg~s$^{-1}$) X-ray sources
with thermal spectrum, but despite many searches no good candidates were
found.  The situation seemed to change in 1996, when the source RX
J1856.5-3754 was discovered \cite{walter96}.  Later studies, however,
demonstrated that the object has a different nature.

 RX J1856.5-3754 and six similar objects (below we will name them The Magnificent seven, or M7 for short) finally were identified as close-by young cooling NSs (see a review in \cite{m7, m72}). For all of them measurements of spin periods are reported (however, for one the result is in doubt). Spin periods are in the range 3-12~s. For several objects period derivative are measured, too \cite{kv09}. In the $P$~--~$\dot P$ diagram it places them between standard radio pulsars and magnetars. However, M7 do not demonstrate either magnetar types of activity, or any radio pulsar or RRAT-like activity \cite{kond}.

M7 are characterized by thermal spectra with a broad spectral feature in most of them \cite{m7}. The origin of this depression is not clear \cite{m72}. Observed emission is mainly due to residual heat and additional heating from the magnetic field decay. 

 Surprisingly, the number of M7-like sources was not increasing since 2001. The best new candidate was identified in 2009 \cite{pires}, however, its nature is not confirmed, and recently a period of 19 msec was reported for this object \cite{pires2}. Population studies (see below) suggest that few ten of objects like these can be identified up to the fluxes an order of magnitude smaller than for known sources. 

\subsection{Magnetars}

 The most popular types of young NSs different from standard radio pulsars are related to the magnetar hypothesis \cite{dt92}.  Now several types of sources are linked to magnetars on different stages of their evolution, but still we have to  start with soft gamma-ray repeaters (SGRs) and anomalous X-ray pulsars (AXPs). 

 The bench mark of the history of magnetars is the discovery of the giant flare of SGR 0526-66 in 1979 \cite{mazets79, vedrenne79}. Now in the McGill on-line catalogue \footnote{http://www.physics.mcgill.ca/$\sim$pulsar/magnetar/main.html} there are two dozens of magnetars, roughly one third of which are candidates, and the numbers of SGRs and AXPs are approximately equal. Difference between these two types of magnetars (here and below we assume that the magnetar hypothesis is true) is very often of historical nature. If an object is discovered as a bursting source --- then it is classified as  a SGR. If an object is discovered as an X-ray source with properties typical for AXPs --- then it is included in the corresponding list. Later very often SGRs are observed as stable X-ray sources, similar to AXPs, and some AXPs demonstrate bursts, similar to so-called weak bursts of
SGRs (see \cite{re11} for a review of bursting activity of magnetars).  

 In a standard picture (which still lacks ultimate confirmation) SGRs and
AXPs are isolated NSs with very strong magnetic fields.  There present day
dipolar fields are $\sim10^{14}$~--~$10^{15}$~G.  They are young objects with
ages $\sim 10^3$~--~$10^5$~yrs.  Due to large dipolar fields they have large
period derivatives, and long (for such young age) spin periods about few
seconds.  
Relatively large stable X-ray luminosity is related to dissipation of the
energy of magnetic fields (currents) in the crust of these NSs.  Strong
bursts are thought to be due to global reconfiguration of the outer magnetic
field structure, which is also linked to processes of field decay in the
crust.


 There are several strong arguments in favour of the magnetar hy\-po\-the\-sis~\footnote{Probably, one of the strongest is related to the existence of strong bursts. Magnetic field energy dissipation is the most conservative mechanism which can explain release of $>10^{40}$~erg in a fraction of a second, and do it repeatedly.}. Let us stop on one of them (see a review and original references in \cite{mer08}).  Thanks to observations aboard the INTEGRAL satellite it became possible to obtain magnetar spectra up to $\sim$~200~keV. They appear to be very hard. I.e., there is a very strong non-thermal component at the energies above $\sim 20$~keV which goes up to 100-200 keV.  There is no unique explanation for this phenomenon in the standard magnetar picture. However, one of reasonable hypothesis  is the following. Thermal photons from the surface are scattered on hot electrons in the magnetosphere. To support the necessary number of high energy electrons it is necessary to have large external magnetic field. Quantitative estimates show that necessary fields fit the standard magnetar hypothesis. 

\subsection{Twilight objects}

 In this subsection we briefly comment on several sources, which can be very important to establish links between different types of NSs and to understand better evolution of compact objects. They can represent transitional phases, or evolved stages.


 We start with the PSR J1846-0258. The object was known as an X-ray pulsar with P=0.326~s inside the SNR Kes 75. The radio pulses are not detected, but clearly it is just due to orientation, as the object has a pulsar wind nebula around. So, it looked like normal young pulsar with estimated magnetic field $B=5 \,
10^{13}$~G. However, in 2008  two papers appeared which reported magnetar-like behavior of this source \cite{ks08, gav08}. 

 The X-ray luminosity of the pulsar increased to the level when it cannot be explained by rotation energy losses and, in addition,  bursts, similar to SGRs and AXPs ones, were detected. 

 This discovery clearly demonstrated that sources can quickly lapse from type to type, and that some internal characteristics are very important.


Another interesting source can be treated as an opposite example, in some
sense.  This is the PSR J1622-4950, detected in a radio survey
\cite{levin10}.  The object has spin period 4.3~s and magnetic field $\sim
3\, 10^{14}$~G.  These values are typical for magnetars, not for radio
pulsar, but the source demonstrate only radio pulsar behavior (however, with
some pecularities like properties of radio spectrum and significant timing
noise).


 The third source related to ``twilight magnetars'' is a recently discovered SGR. The puzzling feature of SGR 0418+572 is its very low dipolar magnetic field. More than one year of observations did not allow to detect spin-down of this, in other respects, normal magnetar with $P=9.1$~s \cite{rea10}. The reported upper limit on the field estimated as $B=3.2 \, 10^{19} \sqrt{P \dot P}$ is $<7.5 \, 10^{12}$~G.  

All three discoveries clearly show that the knowledge of $P$ and $\dot P$ (i.e. external dipolar field) is not enough to conclude about possible appearance (and, hence, classification) of a source. Some parameters that avoid easy detection are crucially important. In the standard picture this role is played by internal (crustal) toroidal field.


The fourth (and the last) object discussed in this subsection is a different one. It got its name --- Calvera
--- as initially it was suspected that it can be related to the Magnificent seven \cite{calvera}. The source showed a thermal spectrum, no traces of other types of activity --- so, it was suspected that it can be a cooling isolated NS. Later observations confirmed the thermal (two-component) spectrum, but the measurements of spin period gave a surprise: 59 msec \cite{zane10}. Upper limit on the period derivative allowed to put an upper limit on the external dipolar field: $<5 \, 10^{10}$~G. Then the object cannot be a relative of the Magnificent seven, but can be an evolved version of CCOs. 

\subsection{X-ray binary systems}

 NSs are product of massive stars evolution. Most of massive stars are members of binary systems. About 10\% of them survive after the first supernova explosion. Then we can expect to find NSs in binary systems. Indeed, there are hundreds of different sources related to them. At the present time all known NSs in binary systems are either accreting (permanently or tran\-sient\-ly), or are radio pulsars. No magnetars activity, and no NSs emitting the residual heat are known in binaries, yet. Very briefly we comment on known types of NS binaries. More detailed information can be found, for example, in \cite{py06} and references therein.

 Most of radio pulsars found in binaries are not young NSs with fields
$\sim 10^{12}$~G.  They are so-called millisecond radio pulsars.  With spin
periods about 10 msec they have fields $\sim 10^9$G and occupy lower left
part of the $P$~--~$\dot P$ diagram.  Millisecond pulsars are old objects. 
They have been spun-up in close binary systems by accretion.  At the same
time their magnetic fields decayed, and the mass was increased. 
When the phase of active accretion was over, they resurrected as radio
pulsars.  Typical companions of millisecond pulsars are old stars: low-mass
stars and white dwarfs.

 Standard radio pulsars also can be found in binary systems. Their com\-pa\-nions can be of different mass, but of course they cannot be significantly evolved. An interesting situation can happen, when a powerful radio pulsar is in pair with a massive star with strong stellar wind. Interaction of relativistic pulsar wind with dense wind of a massive star results in an appearance of a gamma-ray source.

 Finally, a radio pulsar can have as a companion to another NS. Naturally, we expect that in a typical situation we see an older NS as a millisecond radio pulsar, and the younger one can be a dead pulsar. One important exception is the system J0737-3039, where two pulsars were observed. See a review on binary pulsars in \cite{lorimer08}.

 Systems with accreting NSs are divided into two main groups: low-mass X-ray binaries and high-mass X-ray binaries. In the first accretion mostly proceeds via Roche lobe overflow. In the second
--- it can be also due to capture of the stellar wind matter. 

 A NS can appear as an X-ray pulsar of it is accreting, as the falling matter reaches the surface inhomogeneously due to the magnetic field. An important example are millisecond X-ray pulsars which represent the stage prior to the radio millisecond pulsar, i.e. they are NSs in the phase of spin-up.

 Another interesting situation, related to studies of the internal structure of NSs, is  when accretion proceeds transiently. Then, in between two episodes of accretion one can observe a cooling NS. But it is not the same situation that we have with young coolers, emitting their residual heat. Here thermal emission is related to pycnonuclear reactions in the crust of a NS.  See details and references in \cite{ylg05}.

\section{Answers we want to get}

 Why do we study NSs? The first answer is obvious: ``Because they are
beauti\-ful!'' More serious answers can be divided into two parts:
astrophysical and physical.

\subsection{Astrophysical problems}

This part is mainly related to properties and evolution of different sources, to possible relations and differences between them.


The list of questions is inevitably personal, subjective; and I would like to start it with initial properties of NSs. It is of great importance for astrophysical studies to have knowledge about initial distributions of the main NS para\-me\-ters: masses, velocities, spin periods, magnetic fields, etc. In the first place it is important for our understanding of NSs and their evolution. But not only. Initial parameters of compact objects have imprints of supernova explosions, which beget them. Many astrophysical studies of NSs have the final goal in reconstructing some of those distributions of parameters at birth.


Now we know that newborn NSs can have very different sets of parameters. It is crucial to understand what stands behind this difference. All the variety of young NSs described above must find its explanation, if we figure out how diverse combinations of characteristics appear. 


Finally, we observe more or less evolved NSs. We cannot get a clear astrophysical picture for these objects if we do not know evolutionary laws for them. The problem aggravates as many parameters which can be important for the evolution are not directly measured. For example, the angle between spin and magnetic axis, or toroidal magnetic field in a crust. The first of these two can be influential in the magneto-rotational evolution (for example, for radio pulsars), the second --- in thermal evolution and bursting activity of magnetars.  Present evolutionary models in many respects are either over-simplified, or uncertain. 

\subsection{Physical problems}


Problems discussed here are related to the fact that NSs have many extreme properties, and so can serve as labs and instruments to study behaviour of matter under these extreme conditions. Many examples can be mentioned here. Among them electrodynamics in superschwinger magnetic fields, con\-den\-sed states in strong magnetic fields, etc.  Lets us briefly detalize two of them.  

Despite the fact that the history of radio pulsar studies is more than 40 years long, exact mechanisms of radio pulsar spin-down and emission at different wavelengths are not known. This is more a physical than an astrophysical problem.  Observations, of course, help to have some progress in understanding of these phenomenae, still something is missing in the complicated picture of the radio pulsar machinery. 

Many researchers deem that the main physical problem related to NSs is behaviour of matter under high (super-nuclear) density. Indeed, any studies of cold matter with density above few nuclear densities (especially, matter enriched by neutrons) is impossible in the earth laboratories now, and will be impossible in the near future. However, many important issues in the nuclear physics deserves understanding of such behaviour. Numerous research groups allover the world are developing different methods to get some new knowledge in this field. This includes pure theoretical physics studies, extensive numerical modeling, inventing new methods to confront theoretical predictions and observational data, and last, but not the least, new observations to measure NS parameters that can tell us something about properties of their interiors. Below we describe some of them. 

\section{Neutron star interiors and their studies from ``infinity''}

 It is modest to say that astrophysical studies of NS interiors are
``indirect''.  We are dealing with a peculiar situation, when we have to
study inner properties of some object just looking at from a huge distance. 
A NS at several hundreds parsec (typical distance to the closest known
sources of this kind) have the same angular size as a proton, when you look
at it from the distance of several tens of centimeters.  Although, this is the only
way.

 After half-century of discussions and attempts we have several interesting possibilities to learn something about NS structure and characteristics of inner regions. Many measured characteristics of NSs are model-dependent, but some are not. These are mainly temporal parameters: spin period and its variations. Some of them can be used to constrain internal properties of NSs. Already spin period itself is important. Let us remember, that short spin periods were one of the main argument in favour of NS interpretation of newly discovered pulsars: white dwarfs simply could not rotate that rapidly. Observations of very rapidly rotating NSs (with periods $\lesssim $~msec) can provide strong limits on some models of the internal structure. Spin periods are changing on different time scales. Gradual changes can have positive of negative sign in different types of sources. Radio pulsars spin-down, accreting pulsars can
both:  spin-up and spin-down. These data on period variations can be used to put some rough constraints on the structural properties of NSs. Sudden changes of spin frequency of radio pulsars --- so-called gliches --- provide very important information about internal properties of NSs. High frequency oscillations observed in light curves of some  magnetars after strong flares are used to constrain properties of the crust \cite{watts}. Measurements of long-term  timing behaviour of binary radio pulsars can provide direct measurements of moment of inertia, which can be also used to probe the EoS of dense matter.  Temporal characteristics of the gravitational wave signal from coalescences of binary NSs in near future will be also an important source of information about NS matter.

However, now the three main approaches to study NS EoS are: mass determination, radius determination, and studies of thermal evolution of NSs. 

\subsection{Mass determination}

 There are two ways to measure mass of a star. We start with the second --- gravitational redshift. Why with it? --- it is faster to explain. Why the second --- because it is not used, yet. 

 If a redshifted line is identified in a NS spectrum, then we immediately obtain a measure of the ratio $M/R$. This is a unique way to obtain some information about masses of isolated NSs. Unfortunately, now there are no sustainable redshift measurements for NSs. Here we leave this method and come to the only used one.

 Stellar masses are determined if an object is a member of a binary system. 
In the ideal case we can measure orbital velocities of both objects, eccentricity, and orientation of the orbit respect to us. Then, for the mass we have:

\begin{equation}
M_1=(1/2\pi G) (1/\sin ^3 i) K_2^3 P_\mathrm{orb} (1-e^2)^{3/2} (1+M_2/M_1).
\end{equation}  
Here $P_\mathrm{orb}$ --- is the orbital period, $e$ --- eccentricity, $i$ --- is the angle between the normal to the orbit and the line of sight, $K_2$ --- is semi-amplitude of the velocity of the second star. 

Very often not all information is available. Then it is useful to introduce so-called mass function, $f_m$. It can be derived from observations of the second star, and then the mass of the first one is:

\begin{equation}
M_1=f_m (1/\sin ^3 i) (1+M_2/M_1)^2.
\end{equation} 
Obviously, $M_1>f_m$. Then the mass function can be used as a lower limit on the mass. For example, mass function $>3 \, M_\odot$ is an indication that a compact companion is a BH.

There are two main types of systems where NS masses are measured: X-ray  and radio pulsar binaries. Typically, in the first case measurements are less precise. A recent list of mass measurements can be found in \cite{mass2011}.
In the unique case of the binary pulsar PSR J0737-3039 (where both components are NSs observed as radio pulsars) masses of both components are known after just few years of observations with precision $\sim 0.001\, M_\odot$.

It is of particular interest to find most massive NSs, as it allows to eliminate some EoS. For each EoS there is a maximum mass. More massive objects collapse into a black hole.  Thus, a single measurement of a very high mass can lead to a significant progress in our understanding of high density physics. At the moment the highest measured mass is $\sim 2\, M_\odot$ determined for PSR
J1614-2230 \cite{mass21}. Discussion of this important result can be found in \cite{masslatt}. Even higher mass was reported for PSR B1957+20 \cite{masshigh}, but this result is less certain at the moment.

\subsection{Radius determination}

Radius measurements are usually more model dependent than mass mea\-su\-re\-ments. At first, a radius can be estimated for thermally emitting NS using usual formula: $L=4\pi R^2 \sigma T_{\mathrm{eff}}^4$. A clear oversimplification is related to the fact that temperature distribution on the surface can be very inhomogeneous. 
Still, this method is used for different objects, for example, in the case of an isolated cooling NS RX J1856-3754.

 Another approach is used in accreting systems. If an accretion disc is formed, than it is possible to measure redshift at the inner edge of the disc. This is done as follows. If a famous iron line (6.4 keV) is detected in a source, then one can study distortion of the line due to gravitational redshift. Fitting the line it is possible to determine the radius of the disc inner edge (it is mass dependent, however). Then this estimate can be used as an upper limit on the NS radius (see description of the method and some results in \cite{disc} and references therein).

 In some binary systems quasi-regular X-ray bursts appear due to runaway thermonuclear reaction in the accreted layer of hydrogen. Then outer layers are lifted above the NS surface and slowly come back, levitating due to high luminosity. As the first approximation we can take that luminosity is equal to the critical (Eddington) value. Fitting the spectrum with some reasonable model we can obtain an estimate of the radius of the NS. This method is used for quite a long time (see \cite{disc} and references therein) with modifications. One of the main problem is in preparing a good model for the spectrum. Additional difficulties are related to unknown composition of the accreted matter, and to uncertainties in luminosity, which can deviate from the Eddington value.  

 It would be nice to use several different methods to estimate mass and radius in one source. A possibility of this is described in \cite{ozel}, but unfor\-tu\-na\-te\-ly, at the moment this is just a possibility. Despite the fact that several times appeared reports that for some sources it was possible to have mass and radius measurements with small errors, recent studies demonstrate that un\-cer\-tain\-ties are still large \cite{suleimanov, steiner}. 

\section{Thermal evolution of neutron stars}

 NS cooling during first $\sim 10^5$~yrs of evolution is mainly determined by neutrino processes in its interiors. Consequently, observing the surface tem\-pe\-ra\-tu\-re in the case when we can neglect accretion, polar caps heating due to currents in a magnetosphere (positrons going towards NS surface), etc. it is possible to get indirect information about properties of high density matter. 


 There are several reactions working in NS inner regions in which neutrinos are emitted. The most efficient one is the direct URCA process (see a detailed discussion in \cite{hpy2007} and a shorter review in \cite{pethick}). This reaction is allowed if density is above some limit, automatically this means that the process is on if a NS mass is larger then some critical one. If direct URCA is operating inside a NS, then it cools down really fast. Observations demonstrate that there are several NSs of different classes (CCOs, radio pulsar, M7) which are hot at ages $\sim 10^4$~--~$10^6$~yrs. I.e., direct URCA does not set in. This is an important information, which could not be obtained by other matters. 

The example above shows, that indeed observations of cooling NSs can tell us something about internal properties of these objects. Of course, this is just one illustration. Fitting a plethora of observational data on NS thermal properties can give significant support in favour or against a given model of NS structure and EoS. 


 Some of neutrino processes in NSs can be related to neutron and proton superfluidities (see a review in \cite{yakufn}).  Therefore, we can probe the existence of superfluidity in compact objects via observation of their thermal surface emission. The most extreme example was given recently by studies of a CCO Cas A. 

 It was discovered, that the temperature of this source decreased by few percent in several years \cite{ho}. Such a rapid cooling is a signature of a specific stage of NS interiors. Several interpretations are suggested. One of them states that rapid cooling appears due to onset of neutron superfluidity \cite{shternin, page2010}.


NSs in binaries also can provide important information due to their thermal emission. These are compact objects in so-called soft X-ray transients (see, for example, \cite{hpy2007}). NSs in such systems are accreting transiently. Accretion can be switched-off for several years. In this period we can observe surface emission. Suprisingly, thermal radiation is strong, so that it is necessary to involve some additional source of energy. Pycnonuclear reaction can do the job \cite{zdunik}.    Accretion results in sinking of the crust matter into deep, where density is higher. Nuclei appear out of equilibrium, and pycnonuclear reaction stars, heating the compact object. Part of the energy is transported into central parts and emitted by neutrinos, part is emitted from the surface.

Different EoS models can be confronted against observational data using obser\-va\-tions of cooling NSs in soft X-ray transients \cite{ylg05}. This is an important addition to the data on isolated coolers.

\subsection{External layers}

Neutrino emission is determined by processes in NS core. However, at
``infinity'' we see surface emission, which is sensitive to properties
of outer layers of a NS (the proportion in size between a core and a crust
is roughly as between an orange-peel and the rest of the fruit, but the
crust is much less dense than the core, so just $<1$\% of mass is related
to the crust).  This complicates the picture, as it is necessary to take
into account a thermally insulating envelope and an atmosphere which
distorts the spectrum.


NSs (except very young objects, few years old or younger) are nearly
isothermal (having in mind general relativity effects), except a thin outer
layer with significant temperature gradient.  This layer is known as an
en\-ve\-lo\-pe.  As at ``infinity'' we observe the surface temperature,
we have to know how it is related to the temperature inside a NS.  If no
additional heating mechanisms are operating in the crust then the problem
relaxes to heat transport in the envelope.

This question is actively studied, and several models are on the market
(see, for example, \cite{hpy2007, pethick} and references therein). 
Properties of envelopes depend on their chemical composition and magnetic
fields.  Light element envelopes give an effect of higher temperature in NS
young years ($\lesssim 10^5$~yrs), and lower temperature afterwards.  High
constant dipolar fields in an envelope produce the same effect, but
complicated field geometry and field decay  can modify this
conclusion.


 NSs can have geometrically thin atmospheres with typical scale height
$\sim$~cm.  Because of it, an observed spectrum can deviate significantly
from a pure blackbody.  Two limiting cases are represented by hydrogen and
iron outer layers.  In the first case spectrum is relatively featureless,
but strongly deviates from the blackbody which roughly can be described as a
shift towards higher energy (and, of course, slightly down).  This happens
because for higher energy photons ($\sim$~keV in a realistic case of a NS
with $T\gtrsim 10^6$~K) the atmosphere is more transparent, and they are
able to escape to ``infinity'' from deeper layers, where temperature is
higher.  In the case of an iron atmosphere spectrum for $T\sim10^6$~K does
not deviate from the blackbody significantly, but is expected to have many
lines and edges.  Presence of strong magnetic fields brings additional
complications into calculating properties of outgoing emission (see a review
on NS atmospheres in \cite{zavlin}.

 Unknown properties of an atmosphere can result in wrong estimates for the effective temperature and radius of the emitting region. The latter can be illustrated by the case of Cas A. For several years it was puzzling that fits provided very small emitting area, however, no pulsations have been ever detected from this source.  The mystery was solved in 2009 \cite{casa}. With a carbon atmosphere it was possible to obtain realistic values for temperature and radius, consistent with absence of pulsations due to nearly homogeneous temperature distribution.

 It is necessary to mention, that there exists an additional possibility. Instead of an atmosphere a condensate can form on a surface if temperature is low or/and if the magnetic field is high \cite{condense}. In this case the spectrum also can deviate from a blackbody, but can mimic it in a limited energy range.

\subsection{Additional heating, magnetic field decay}

 Several different mechanisms of additional NS heating have been discussed
in the literature.  We focus just on one of them: magnetic field decay. 
This mechanism is a standard element of magnetar physics.  If it is
important in NSs with standard ($\sim 10^{12}$ G) or small ($\lesssim
10^{10}$~G) magnetic fields is not so clear.  But for high-field NSs
additional heating due to field decay fits the data quite well \cite{link}.

\begin{figure}[h]
 \centerline{\includegraphics[width=0.8\textwidth]{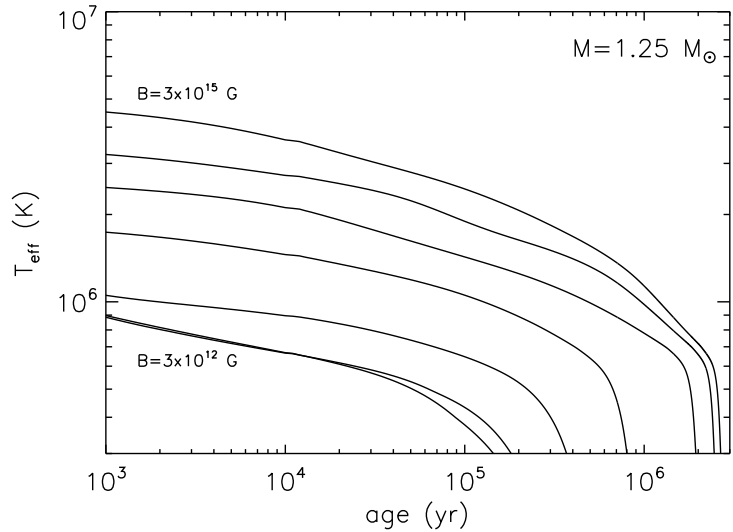}}
  \caption{Cooling curves of a $M=1.25 M_\odot$ NS with different values of the initial magnetic field strength. From bottom to top: $B_0= 3 \times 10^{12}, 10^{13},  3 \times 10^{13}, 10^{14}, 3 \times 10^{14}, 10^{15},$ and $3 \times 10^{15}$ G. From \cite{ppmpb2010}.}
\label{fig:b}
\end{figure}

\begin{figure}[h]
 \centerline{\includegraphics[width=0.8\textwidth]{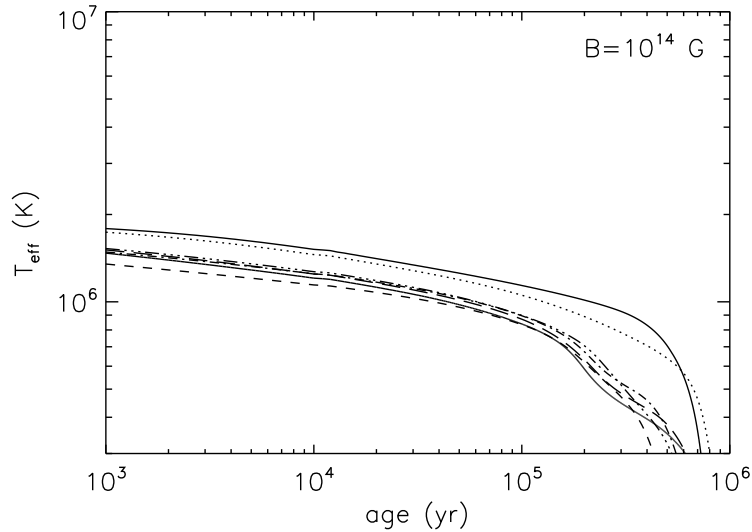}}
  \caption{Cooling curves with $B_0=10^{14}$~G  and different masses: $M=1.10$ (top solid line), 1.25 (dots), 1.32 (dashes), 1.40 (dash-dot), 1.48 (dash-triple dot), 1.60 (long dashes), and 1.70 $M_\odot$ (bottom solid line). From \cite{ppmpb2010}.}
\label{fig:m}
\end{figure}

In Figs.~\ref{fig:b}, \ref{fig:m} we compare cooling curves obtained in the framework of decaying magnetic field. It is obvious from the figure that for high initial fields ($\sim 10^{14}$~G) the effect of additional heating dominates different effects related to conditions in the interiors (which strongly depend on NS mass).  

\begin{figure}[h]
 \centerline{\includegraphics[width=0.8\textwidth]{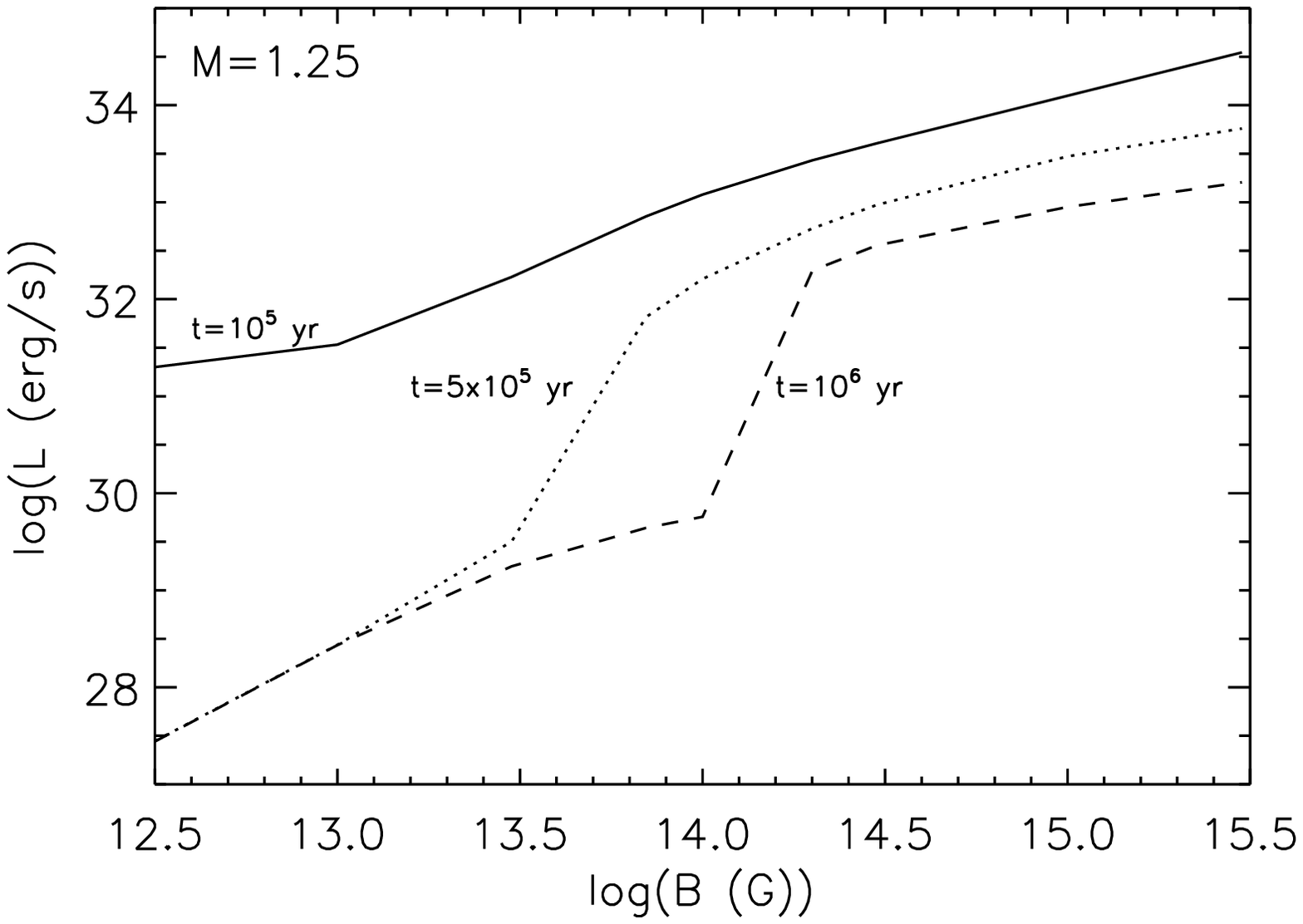}}
  \caption{Luminosity as a function of the initial magnetic field strength for a $M=1.25 M_\odot$ NS at different ages ($10^5$, $5\times10^5$, and $10^6$ years). From \cite{ppmpb2010}.}
\label{fig:l}
\end{figure}

Magnetic field decay makes the NS evolution much more interesting and complicated as several parameters, otherwise independent, become strongly coupled. Magneto-rotational and thermal evolution should be considered in one framework. Let us add evolution of bursting activity and its possible influence on period and temperature changes, and we obtain a very com\-pli\-ca\-ted picture self-consistent description of which is still in future.

\section{Population studies of neutron stars}

In astrophysics we typically can observe just a tip of an iceberg --- only the brighter sources of some given type. In addition, only some specific evolutionary stages for a given type of objects can have detectable mani\-fes\-ta\-tions. Then we need some special approach to deal with such a situation. Such method exists and it is called population synthesis \cite{ufn}.

Population synthesis is realized as a numerical method. Several modi\-fi\-ca\-tions are known, but in general it works as follows. A synthetic population of sources with given distribution of initial parameters is formed. Then it is evolved using prescribed evolutionary laws. Observable parameters are cal\-cu\-la\-ted during this evolution, and statistics is collected. Finally, after normalization a synthetic population is confronted with an observed one (if any exists, otherwise results of population synthesis are predictions for future ob\-ser\-va\-tions). It allows to check the correctness of model assumptions (initial conditions, evolutionary laws, and normalization). 

NSs of different types are actively studied by the population synthesis calculations (see a review in \cite{ufn}). For obvious reasons radio pulsars and NSs in binaries are the most popular objects in these studies. Some types of sources (CCOs, RRATs) were not studied by population synthesis method, yet. At the present time there are many uncertainties in NS evolution, that is why researchers often limit themselves with just one population of sources. This is a willy-nilly situation, but it is necessary to move towards a unified picture, in which all types of NSs and their activities are described in one framework \cite{kaspi10}. This will require efforts in all types of research: observations, theory, modeling of evolution. 


\section*{Conclusions}

 It seems that NS astrophysics will have bright future at least in the following 10-15 years. There are two reasons for such an optimistic conclusion. At first, there are many really interesting well-posed unsolved problems. On the other hand, several new observational facilities of the top class will start their operation in the following years. These included upgraded gravitational wave detectors, large radio telescopes, new X-ray observatories, and large optical telescopes.  We should not forget about new accelerator projects, such as FAIR and NICA, that will also contribute to our knowledge of high density physics.

\bigskip

{\bf Acknowledgements}\\
I would like to thank the Organizers for the wonderful school and for support of my participation in it. It was a pleasure to see many talented, motivated and educated students. I thank for hospitality and support the University of Padova, where this paper was written during my visit, and Prof. Turolla for many interesting discussions.

The work was supported by RFBR grants 10-02-00599 and 12-02-00186,  and by the Federal program for scientific and teaching staff (contract 02.\-740.\-11.\-0575).

\end{document}